\documentclass[amsmath,amssymb,aps,prb,nofootinbib]{revtex4-2}
\usepackage{stix2}
\usepackage{graphicx}% Include figure files
\usepackage{natbib}
\usepackage[colorlinks=true,linkcolor=blue,citecolor=blue,urlcolor=blue]{hyperref}
\usepackage{xcolor}

\begin{document}
\title{Spheres and fibres in turbulent flows at various Reynolds numbers}
\author{Ianto Cannon$^1$}
\author{Stefano Olivieri$^{1,2}$}
\author{Marco E. Rosti$^1$}
\email{marco.rosti@oist.jp}
\affiliation{$^1$Complex Fluids and Flows Unit, Okinawa Institute of Science and Technology Graduate University, 1919-1 Tancha, Onna-son, Okinawa 904-0495, Japan \\
$^2$ Department of Aerospace Engineering, Universidad Carlos III de Madrid, Avda. de la Universidad, 30. 28911 Leganés, Spain}

\date{\today}
\begin{abstract}
We perform fully coupled numerical simulations using immersed boundary methods of finite-size spheres and fibres suspended in a turbulent flow for a range of Taylor Reynolds numbers $12.8<Re_\lambda<442$ and solid mass fractions $0\leq M\leq1$. Both spheres and fibres reduce the turbulence intensity with respect to the single-phase flow at all Reynolds numbers, with fibres causing a more significant reduction than the spheres. {The particles' effect on} the anomalous dissipation tends to vanish as $Re \to \infty$. A scale-by-scale analysis shows that both particle shapes provide a ``spectral shortcut'' to the flow, but the shortcut extends further into the dissipative range in the case of fibres. Multifractal spectra of the near-particle dissipation show that spheres enhance dissipation in two-dimensional sheets, and fibres enhance the dissipation in structures with a dimension greater than one and less than two. In addition, we show that spheres suppress vortical flow structures, whereas fibres produce structures which completely overcome the turbulent vortex stretching behaviour in their vicinity.
\end{abstract}
\maketitle

\begin{figure}
\begin{minipage}{1.07\textwidth}
\raisebox{-0.5\height}{
	\includegraphics[width=0.85\textwidth]{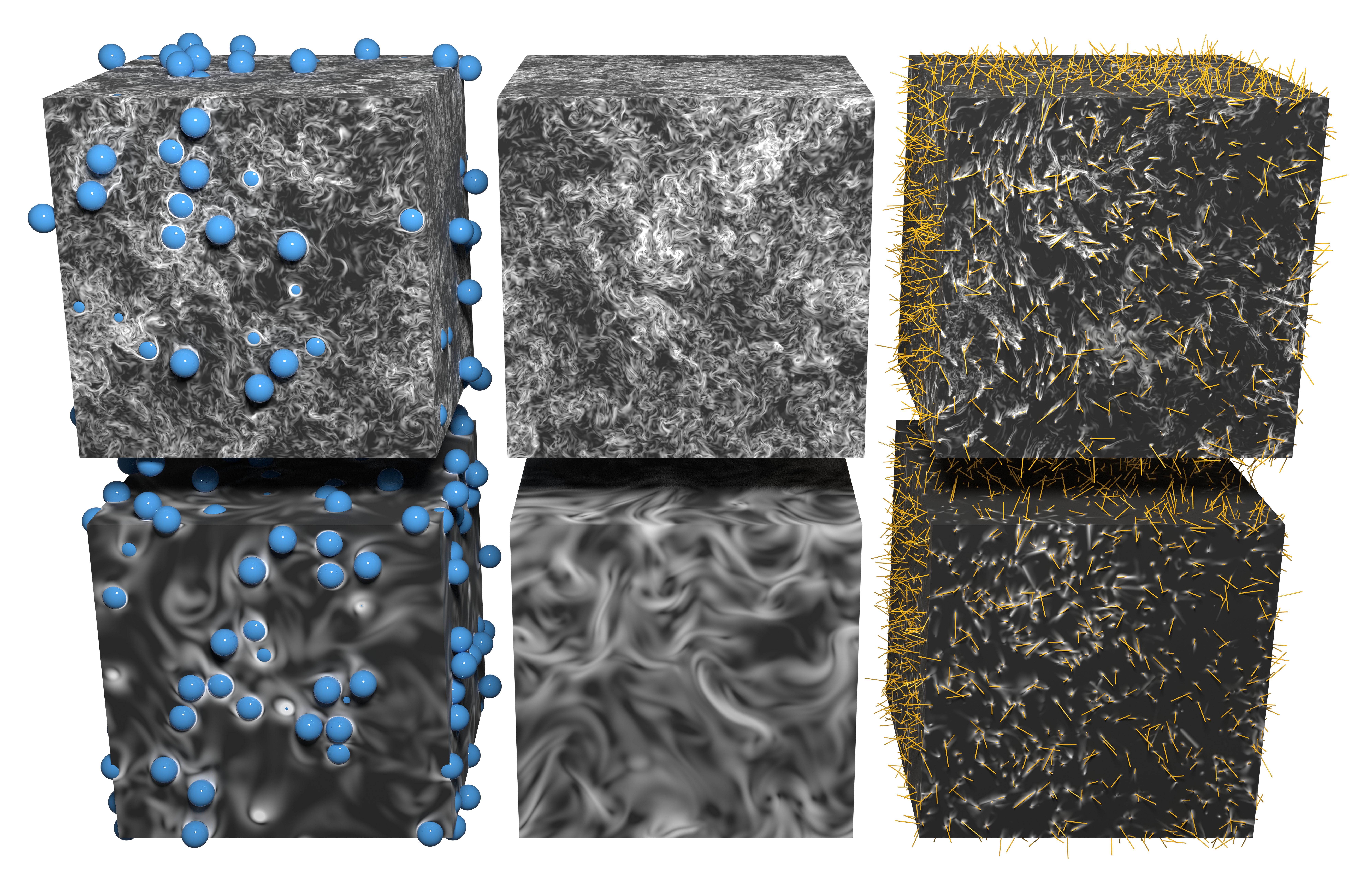}
}
\raisebox{-0.5\height}{
	\includegraphics[width=0.05\textwidth,trim={.7cm 0 0 0}]{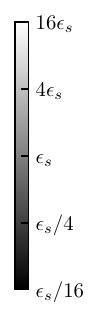}
}
\end{minipage}
\caption{Images of the simulated domain. Left column: flows with fixed spheres. Middle column: single phase flows. Right column: flows with fixed fibres. Flows on the top row have $Re_{ABC}=894$. Flows on the bottom row have $Re_{ABC}=55.9$. At the boundary of each domain, we show the dissipation on a logarithmic scale, where $\epsilon_s$ is the mean dissipation of the single-phase flow at $Re_{ABC}=894$.}
 \label{fig:snapshots}
\end{figure}

\section{Introduction}
\label{sec:intro}
Particle-laden turbulent flows abound in our environment; see, for example, volcanic ash clouds, sandstorms, and ocean microplastics. In these cases, the particles are seen in a range of shapes, and the intensity of the carrier turbulent flow can significantly vary. In this context, the present study delves into the interaction between particles and turbulent flows, focusing on two distinct particle shapes, spheres and fibres, and exploring turbulent flows across a large range of Reynolds numbers.

The interaction of particles with turbulent flows has garnered significant research attention since~\citet{richardson_diffusion_1927} used balloons to track eddies in the Earth's atmosphere in 1927. Light spherical particles as tracers are now a ubiquitous experimental tool~\cite{westerweel_particle_2013}, and even fibre-shaped tracers have been employed~\cite{brizzolara_fiber_2021}. For heavier particles, more complex motion is seen, such as inertial clustering~\cite{monchaux_measuring_2012}, preferential sampling~\cite{goto_sweep-stick_2008}, {and enhancement of the mean flow~\cite{chiarini_anisotropic_2024}}. Numerical simulations in this field have been a valuable tool, as results can be readily processed to study length-scale dependent statistics such as radial distribution functions~\cite{ireland_effect_2016}, and histograms of Vorono\"i area~\cite{monchaux_measuring_2012}. Most of such particle-laden turbulent studies make the one-way coupling assumption, whereby the particles are affected by the fluid motion, but the fluid does not feel any effect of the particles. When the total mass fraction or volume fraction of particles in the system increases, the one-way coupling assumption becomes invalid, and we must model the back-reaction effect of the particles on the fluid. This opens up a zoo of new phenomena, including drag reduction~\cite{zhao_turbulence_2010}, cluster-induced turbulence~\cite{capecelatro_fluidparticle_2015}, and new scalings in the energy spectrum~\cite{olivieri_effect_2022}. A recent review by~\citet{brandt_particle-laden_2021} has pointed out a gap in the current understanding of particles in turbulence, which lies between small heavy particles and large weakly buoyant particles. In addition, most real-world particle-laden flows involve particles with a high degree of anisotropy, while research in particle-laden turbulence has overwhelmingly focused on spherical particles. 

This article addresses the gaps by studying a range of particle mass fractions $M$, ranging from single-phase ($M=0$) to fixed particles ($M=1$). We make simulations using a fully coupled approach to elucidate how particles modulate turbulence. Furthermore, we vary the turbulence intensity, allowing us to ask the question \emph{at what Reynolds numbers ($Re$) does turbulence modulation emerge?} And \emph{does the modulation effect persist as $Re\to\infty$?} Crucially, we connect our results to real-world flows by investigating isotropically-shaped particles (spheres) and anisotropically-shaped particles (fibres). The spheres have a single characteristic length, given by their diameter, whereas the fibres have two: their length and thickness. This naturally allows us to ask \emph{how do the particle's characteristic lengths impact the scales of the turbulent flow?}

A number of works have investigated particle shape; see~\citet{voth_anisotropic_2017} for a review of the behaviour of anisotropic particles in turbulence, including oblate spheroids, prolate spheroids and fibres. In 1932 \citet{wadell_volume_1932} defined the sphericity of a particle as its surface area divided by the area of a sphere with equivalent volume; Wadell used sphericity to classify quartz rocks according to their shape. More recently, \citet{zhao_rotation_2015} made direct numerical simulations of oblate and prolate spheroids in a channel flow and found that away from the channel walls, prolate spheroids tend to rotate about their symmetry axis (spinning), whereas oblate particles rotate about an axis perpendicular to their symmetry axis (tumbling). \citet{ardekani_drag_2017} showed that oblate spheroids can reduce drag in a turbulent channel by aligning their major axes parallel to the wall. \citet{yousefi_modulation_2020} simulated spheres and oblate spheroids in a shear flow and found that the rotation of the spheroids can enhance the kinetic energy of the flow.
At the other shape extreme from oblate spheroids, we have fibres, which are long and thin. In this article, we choose to study rigid fibres (also known as rods) as they are a simple example of a particle with high anisotropy. Concerning fibres, \citet{paschkewitz_numerical_2004,gillissen_fibre-induced_2008} and others showed that rigid fibres align with vorticity in channel flows, dissipating the vortex structures, and drag reductions up to 26\% have been measured~\cite{paschkewitz_numerical_2004}. In this article, we wish to investigate how the shapes and length scales of particles interact with turbulence, so we choose the triperiodic flow geometry. This geometry avoids the effects of walls and other large structures on the flow, enabling us to focus on the emergent length scales of the flow.

A few works have investigated the effect of turbulence intensity on particle-laden flows. \citet{lucci_modulation_2010} simulated spheres of various diameters $c$ in decaying isotropic turbulence and found that spheres reduce the fluid kinetic energy and enhance dissipation when $c>\eta$, where $\eta$ is the Kolmogorov length scale. In this article, we also vary the ratio $c/\eta$. However, unlike~\cite{lucci_modulation_2010}, we do it by varying the fluid viscosity, not the particle size. This enables us to keep the particle volume, particle surface area and number of particles constant across our cases. \citet{oka_attenuation_2022-1} studied sphere diameters in the range $7.8<c/\eta<64$, in turbulent flows with $Re_\lambda\approx94$ and showed that vortex shedding and turbulence attenuation occur when $c\gtrsim\lambda\rho/\tilde\rho$, where $\lambda$ is the Taylor length-scale, $\rho$ and $\tilde\rho$ are the fluid and solid densities, respectively. \citet{shen_turbulence_2022} investigated the effect of solid-fluid density ratio for flows with $38<Re_\lambda<74$, and $8.8<c/\eta<18$, showing that higher density spheres cause {an increase in the normalised dissipation rate and a greater} turbulence attenuation.
\citet{peng_parameterization_2023} parametrized the attenuation of turbulent kinetic energy by spheres with $7.1<c/\eta<15$ and $41<Re_\lambda<63$. They found that the particle mass fraction is indeed a strong indicator of attenuation, and there is negligible dependence on the Taylor Reynolds number $Re_\lambda$ for this range. Compared to the above works, we choose a wider range of turbulence intensities, such that $11.7<c/\eta<125$ and $12.8<Re_\lambda<442$, and we extend the study to include non-spherical particles. 

This article is structured as follows: In the following section, we describe the numerical methods and the parameters used in our study. In section~\ref{sec:results}, we present and discuss the results of our simulations, and section~\ref{sec:concl} concludes with a summary and outlook on future research.

\section{Methods and setup}
We tackle the problem using large direct numerical simulations on an Eulerian grid of $1024^3$ points with periodic boundaries in all three directions. To obtain the fluid velocity $u_i$ and pressure $p$, we solve the incompressible Navier-Stokes equations for a Newtonian fluid with kinematic viscosity $\nu$ and density $\rho$,
\begin{align}
	\partial_t u_i + \partial_j (u_i u_j)  &= \nu \partial_{jj} u_i -\partial_i p/\rho + f^{ABC}_i + f^{sf}_i,\label{eqNS}\\
	\partial_j u_j &= 0,\label{eqIncomp}
\end{align}
where indices $i,j\in\{1,2,3\}$ denote the Cartesian components of a vector, and repeated indices are implicitly summed over. The turbulent flow is sustained by an ABC forcing \citep{libin_long_1990,podvigina_non-linear_1994}, which is made of sinusoids with a wavelength $2\pi L$ equal to the domain size,
\begin{equation}
\begin{split}
	f^{ABC}_1= C\sin({x_3}/{L}) + C\cos({x_2}/{L}),\\
	f^{ABC}_2= C\sin({x_1}/{L}) + C\cos({x_3}/{L}),\\
	f^{ABC}_3= C\sin({x_2}/{L}) + C\cos({x_1}/{L}).
\end{split}
\label{eqABC}
\end{equation}
The amplitude $C$ of the forcing is used to define the forcing Reynolds number ${Re_{ABC}\equiv C^{1/2}L^{3/2}/\nu}$. To discern the effect of increasing turbulence intensity, we conduct a number of simulations with various values of $Re_{ABC}$, given in table~\ref{tabSinglePhCases}.

When the single-phase flows reach a statistically steady state, we add the solid particles at random (non-overlapping) locations and orientations in the domain. We allow the flow to reach a statistically steady state again, and measure the statistics presented in section~\ref{sec:results}, which were averaged in time over multiple large-eddy turnover times $\tau_f\equiv2\pi L/u_{{rms}}$, where $u_{{rms}}$ is the root mean square of the fluid velocity. This procedure is repeated for every solid mass fraction investigated, defined as $M\equiv m_{s}/(m_{s}+m_{f})$ where $m_{s}$ and $m_{f}$ are the total mass of solid and fluid. The single-phase cases have $M=0$, and the cases with fixed particles have $M=1$. The solid phase is two-way coupled to the fluid using the immersed boundary method, and the back-reaction of the particles on the fluid is enforced by $\mathbf{f^{sf}}$ in equation~\ref{eqNS}. As can be seen in figure~\ref{fig:snapshots}, we simulate two types of particles, spheres and fibres; to isolate the effect of particle isotropy on the flow, we choose the fibres and spheres to have the same size: the spheres have diameter $c$, and the fibres have length $c$. {We choose ${c=L/2}$, which lies inside the inertial range of scales for all of our cases.}

We use the in-house Fortran code \emph{Fujin} to solve the flow numerically. Time integration is carried out using the second-order Adams-Bashforth method, and incompressibility (equation~\ref{eqIncomp}) is enforced in a pressure correction step~\cite{kim_application_1985}, which uses the fast Fourier transform. Variables are defined on a staggered grid; velocities and forces are defined at the cell faces, while pressure is defined at the cell centres. Second-order finite differences are used for all spatial gradients. See \url{https://groups.oist.jp/cffu/code} for validations of the code.

\begin{table}
\centering
\begin{tabular}{ccc|ccc}
marker & $Re_{ABC}$ & $M$ & $\eta/L$ & $Re_\lambda$ & $\epsilon \mathcal{L}/u_{{rms}}^3$ \\[3pt]
%    \hline%    \hline
$\pentagonblack$ & $\mathbf{894}$ &\textbf{0.0} &{$\mathbf{4.06\times10^{-3}}$}       &{\textbf{433}}&    0.399     \\
$\blacksquare  $ & $ 447 $        &\textbf{0.0} & $6.31 \times 10^{-3}$                     &      308           &{\textbf{0.382}}\\
$\blacktriangle$ & $ 224 $        &\textbf{0.0} & $1.11 \times 10^{-2}$                     &      204           &    0.400     \\
$\blacklozenge $ & $ 112 $        &\textbf{0.0} & $1.95 \times 10^{-2}$                     &      116           &{\textbf{0.473}}\\
$\mdlgblkcircle$ & $\mathbf{55.9}$&\textbf{0.0} &{$\mathbf{2.99\times10^{-2}}$}       &{\textbf{101}}&    0.428     \\
  \end{tabular}\\[9pt]
\caption{Single-phase flows. $Re_{ABC}$ is the forcing Reynolds number, and $M$ is the solid mass fraction. We measure the Kolmogorov length scale $\eta\equiv\nu^{3/4}/\epsilon^{1/4}$, the Taylor Reynolds number $Re_\lambda$, and the dissipation $\epsilon$, which has been normalised using the integral length scale $\mathcal{L}$ and the root mean square velocity $u_{{rms}}$ of the fluid. The largest and smallest values of each parameter are shown in bold.}
\label{tabSinglePhCases}
\end{table}

\subsection{Motion of the spheres}
The sphere motion and forces are modelled using an Eulerian-based immersed boundary method developed by~\citet{hori_eulerian-based_2022}. The velocity $\mathbf{U}$ and rotation rate $\mathbf{\Omega}$ of each sphere are found by integrating the Newton-Euler equations in time
\begin{align}
	m\partial_t U_i &= \oiint_{S} \left(\rho\nu(\partial_i u_j + \partial_j u_i)-p\delta_{ij}\right)n_jdS-F^{col}n_i,	\\
	I\partial_t \Omega_i&= \oiint_{S} {\varepsilon_{ijk}}\frac{c}{2}n_j\rho\nu(\partial_k u_l + \partial_l u_k)n_l dS,
\end{align}
where $\delta_{ij}$ is the Kronecker delta and ${\varepsilon_{ijk}}$ is the Levi-Civita symbol, $S$ is the surface of the sphere, and $\mathbf{n}$ is its normal. ${m=\tilde\rho\pi c^3/6}$ and ${I=mc^3/20}$ are the mass and moment of inertia of the sphere with diameter $c$ and density $\tilde\rho$. A soft-sphere collision force $F^{col}\mathbf n$ is applied in the radial direction when spheres overlap \cite{hori_eulerian-based_2022}. 

Table~\ref{tabSphCases} shows our choice of parameters for the sphere-laden flows. In all cases, we use 300 spheres, {giving the solid phase a volume fraction of $0.079$}. The characteristic time of the spheres is $\tau_s=\tilde\rho c^2/(18\rho\nu)$, from which we can define the Stokes number $St\equiv\tau_s/\tau_f$ of our flows. The particle Reynolds number of the spheres is $Re_p= c\sqrt{\langle\mathbf{\Delta u}_n\cdot\mathbf{\Delta u}_n\rangle_n}/\nu$, where $\mathbf{\Delta u}_n$ is the difference between the velocity of the $n$th particle and the local fluid velocity, averaged in a ball of diameter $2c$ centred on the sphere. The angled brackets $\langle\rangle_n$ denote an average over all spheres in the simulation.

\begin{table}
\centering
  \begin{tabular}{cccc|ccccc}
marker & $Re_{ABC}$ & $M$ & $\tilde\rho/\rho$ & $\eta/L$ & $Re_\lambda$ & $\epsilon \mathcal{L}/u_{{rms}}^3$ & $St$ & $Re_p$ \\[3pt]
\color[HTML]{023858}$\pentagonblack$ &$\mathbf{894}$&{\textbf{0.1}}&\textbf{1.29}      &{$\mathbf{4.09\times 10^{-3}}$}&{\textbf{431}}&        0.397          &             7.4   &          618             \\  %  iM0p10
\color[HTML]{045a8d}$\pentagonblack$ &$\mathbf{894}$&          0.3       &         4.99      & $4.18 \times 10^{-3}$               &          397       &{\textbf{0.395}} &            26.9   &          857             \\  %  iM0p30
\color[HTML]{0570b0}$\pentagonblack$ &$\mathbf{894}$&          0.6       &         17.4      & $4.09\times 10^{-3}$                &          346       &        0.507          &            89.1   &         1120             \\  %  iM0p60
\color[HTML]{3690c0}$\pentagonblack$ &$\mathbf{894}$&          0.9       &          105      & $4.18 \times 10^{-3}$               &          280       &        0.625          &             471   &         1180 						 \\  %  iM0p90
\color[HTML]{a6bddb}$\pentagonblack$ &$\mathbf{894}$& \textbf{1.0}       &$\boldsymbol\infty$& $4.29 \times 10^{-3}$               &          247       &        0.708          &$\boldsymbol\infty$& \textbf{1190} 					 \\  %  iM1p00
\color[HTML]{a6bddb}$\blacksquare  $  &  $ 447$     & \textbf{1.0}       &$\boldsymbol\infty$& $7.31 \times 10^{-3}$               &          161       &        0.786          &$\boldsymbol\infty$&          559             \\  %  inuP005M1
\color[HTML]{023858}$\blacktriangle$  &  $ 224$     &{\textbf{0.1}}&\textbf{1.29}      & $1.13 \times 10^{-2}$               &          219       &        0.374          &{\textbf{1.91}}&        142             \\  %  inuP01MP1
\color[HTML]{045a8d}$\blacktriangle$  &  $ 224$     &          0.3       &         4.99      & $1.16 \times 10^{-2}$               &          181       &        0.477          &            6.49   &          198             \\  %  inuP01MP3
\color[HTML]{0570b0}$\blacktriangle$  &  $ 224$     &          0.6       &         17.4      & $1.18 \times 10^{-2}$               &          157       &        0.586          &            20.9   &          243             \\  %  inuP01MP6
\color[HTML]{3690c0}$\blacktriangle$  &  $ 224$     &          0.9       &          105      & $1.19 \times 10^{-2}$               &          123       &        0.751          &             109   &          266             \\  %  inuP01MP9
\color[HTML]{a6bddb}$\blacktriangle$  &  $ 224$     & \textbf{1.0}       &$\boldsymbol\infty$& $1.26 \times 10^{-2}$               &          101       &        0.936          &$\boldsymbol\infty$&          260             \\  %  inuP01M1
\color[HTML]{a6bddb}$\blacklozenge $  &  $ 112$     & \textbf{1.0}       &$\boldsymbol\infty$& $2.14 \times 10^{-2}$               &         61.5       &         1.19          &$\boldsymbol\infty$&          117             \\  %  inuP02M1
\color[HTML]{a6bddb}$\mdlgblkcircle$&$\mathbf{55.9}$& \textbf{1.0}       &$\boldsymbol\infty$&{$\mathbf{3.60\times 10^{-2}}$}&{\textbf{34.  9}}&{\textbf{1.81}}&$\boldsymbol\infty$&{\textbf{44.6}}    \\  %  inuP04fixed
\end{tabular}\\[9pt]
\caption{Sphere-laden flows. We set the solid-fluid density ratio $\tilde\rho/\rho$ to obtain a range of solid mass fractions $M$. Bulk statistics for the particles are the Stokes number $St$ and the particle Reynolds number $Re_p$. The largest and smallest values of each parameter are shown in bold.}
\label{tabSphCases}
\end{table}

\subsection{Motion of the fibres}
For the motion and coupling of the fibres, we also use an immersed boundary method, but this one is Lagrangian and solves the Euler-Bernoulli equation for the position $\mathbf{X}$ of the beam with coordinate $l$ along its length~\cite{huang_simulation_2007,alizad_banaei_numerical_2020,olivieri_dispersed_2020}
\begin{equation}
	\frac{\pi}{4} d^2(\tilde\rho-\rho) \partial^2_{t}X_i=\partial_l (T\partial_l X_i) + \gamma \partial^4_{l}X_i-F^{fs}_i+F^{col}_i,
	\label{eq:eulerB}
\end{equation}
where $T$ is the tension, enforcing the inextensibility condition;
\begin{equation}
	\partial_{t}X_i\partial_{t}X_i=1.
\end{equation}
The volumetric density of the fibre is $\tilde\rho$, and its stiffness is $\gamma$. Note that the fluid density $\rho$ in equation~\ref{eq:eulerB} cancels the inertia of the fluid in the fictitious domain inside the fibre~\cite{yu_dlmfd_2005}. To exclude deformation effects in our study, we choose a stiffness which limits the fibre deformation below 1\%. The collision force $\mathbf{F^{col}}$ is the minimal collision model by~\citet{snook_vorticity_2012}. The fluid-solid coupling force $\mathbf{F^{fs}}$ enacts the non-slip condition at the particle surface, and an equal and opposite force ($\mathbf{f^{sf}}$ in equation~\ref{eqNS}) acts on the fluid. The spreading kernel onto the Eulerian grid has a width of three grid spaces, giving the fibre diameter $d=3\Delta x$ in units of the Eulerian grid spacing~{\cite[equation 14]{pinelli_immersed-boundary_2010}}.
{The fibre diameter $0.4\eta<d<5\eta$ is on the order of the Kolmogorov length $\eta$ in all cases, i.e., it is smaller than the turbulent eddies in the energy-containing range of the flow. This allows us to {consider} the fibres as infinitesimally thin objects with a high degree of anisotropy.}
Table~\ref{tabFibCases} shows our choice of parameters for the fibre-laden flows. {In all cases, we use $10^4$ fibres, giving the solid phase a volume fraction of around $5.4\times10^{-3}$}. The characteristic time of the fibres is calculated using a formulation which takes their aspect ratio $\beta\equiv c/d$ into account~\cite{shaik_kinematics_2023},
\begin{equation}
	\tau_s=\frac{\tilde\rho d^2}{18\rho\nu} \frac{\beta\ln \left(\beta+\sqrt{\beta^2-1}\right)}{\sqrt{\beta^2-1}},
\end{equation}
from which we can define the Stokes number $St\equiv\tau_s/\tau_f$ of our flows. The particle Reynolds number of the fibres is $Re_p= d\sqrt{\langle\mathbf{\Delta u}_n\cdot\mathbf{\Delta u}_n\rangle_n}/\nu$, where $\mathbf{\Delta u}_n$ is the difference between the velocity of the midpoint of the $n$th fibre and the local fluid velocity, averaged in a ball of diameter $2c$ centred on the fibre's midpoint. The angled brackets $\langle\rangle_n$ denote an average over all fibres in the simulation.

\begin{table}
\centering
  \begin{tabular}{cccc|ccccc}
marker & $Re_{ABC}$ & $M$ & $\tilde\rho/\rho$ & $\eta/L$ & $Re_\lambda$ & $\epsilon \mathcal{L}/u_{{rms}}^3$ & $St$ & $Re_p$ \\[3pt]
\color[HTML]{662506}$\pentagonblack$ & $\mathbf{894}$&          0.2     &         47.2      & $4.09 \times 10^{-3}$      &          422 &{\textbf{0.425}}&          1.45        &         34.9      \\  %  sM0p20
\color[HTML]{993404}$\pentagonblack$ & $\mathbf{894}$&          0.3     &         81.8      &$\mathbf{4.02\times10^{-3}}$& \textbf{442} &        0.432       &           2.6        &         41.2      \\  %  sM0p30
\color[HTML]{cc4c02}$\pentagonblack$ & $\mathbf{894}$&          0.6     &          279      & $4.17 \times 10^{-3}$      &          340 &        0.713       &          7.52        &         46.4      \\  %  sM0p60
\color[HTML]{ec7014}$\pentagonblack$ & $\mathbf{894}$&          0.9     &         1690      & $4.26 \times 10^{-3}$      &          223 &         1.15       &          36.1        &         47.2      \\  %  sM0p90
\color[HTML]{fec44f}$\pentagonblack$ & $\mathbf{894}$& \textbf{1.0}     &$\boldsymbol\infty$& $4.40 \times 10^{-3}$      &          201 &         1.29       &   $\boldsymbol\infty$&         46.0      \\  %  sRe400_ABC_N1e4_fixed
\color[HTML]{fec44f}$\blacksquare$   &    $ 447$     & \textbf{1.0}     &$\boldsymbol\infty$& $7.44 \times 10^{-3}$      &          115 &         1.78       &   $\boldsymbol\infty$&         20.6      \\  %  sRe200_ABC_N1e4_fixed
\color[HTML]{662506}$\blacktriangle$ &    $ 224$     &{\textbf{0.1}}&         21.3    & $1.16 \times 10^{-2}$      &          192 &        0.434       & \textbf{0.155}       &{\textbf{8.31}}\\  %  snuP01MP1
\color[HTML]{993404}$\blacktriangle$ &    $ 224$     &          0.3     &         81.8      & $1.16 \times 10^{-2}$      &          196 &        0.465       &         0.605        &          9.9      \\  %  snuP01MP3
\color[HTML]{cc4c02}$\blacktriangle$ &    $ 224$     &          0.6     &          279      & $1.19 \times 10^{-2}$      &          167 &         0.63       &          1.85        &         11.2      \\  %  snuP01MP6
\color[HTML]{ec7014}$\blacktriangle$ &    $ 224$     &          0.9     &         1690      & $1.23 \times 10^{-2}$      &         72.5 &         2.25       &          7.16        &         9.11      \\  %  snuP01MP9
\color[HTML]{fec44f}$\blacktriangle$ &    $ 224$     & \textbf{1.0}     &$\boldsymbol\infty$& $1.29 \times 10^{-2}$      &         54.9 &         3.17       &   $\boldsymbol\infty$&         8.22      \\  %  snuP01M1
\color[HTML]{fec44f}$\blacklozenge$  &    $ 112$     & \textbf{1.0}     &$\boldsymbol\infty$& $2.33 \times 10^{-2}$      &         28.8 &         4.68       &   $\boldsymbol\infty$&         3.32      \\  %  sRe50_ABC_N1e4_fixed
\color[HTML]{fec44f}$\mdlgblkcircle$ &$\mathbf{55.9}$& \textbf{1.0}     &$\boldsymbol\infty$&$\mathbf{4.27\times10^{-2}}$& \textbf{12.8}&\textbf{8.60}       &   $\boldsymbol\infty$& \textbf{1.51}     \\  %  sRe25_ABC_N1e4_fixed
\end{tabular}\\[9pt]
\caption{Fibre-laden flows. We set the solid-fluid density ratio $\tilde\rho/\rho$ to obtain a range of solid mass fractions $M$. Bulk statistics for the particles are the Stokes number and particle Reynolds number. The largest and smallest values of each parameter are shown in bold.}
\label{tabFibCases}
\end{table}

\section{Results and discussion}
\label{sec:results}

\subsection{Bulk statistics}
\begin{figure}
\centering
\includegraphics[width=\textwidth]{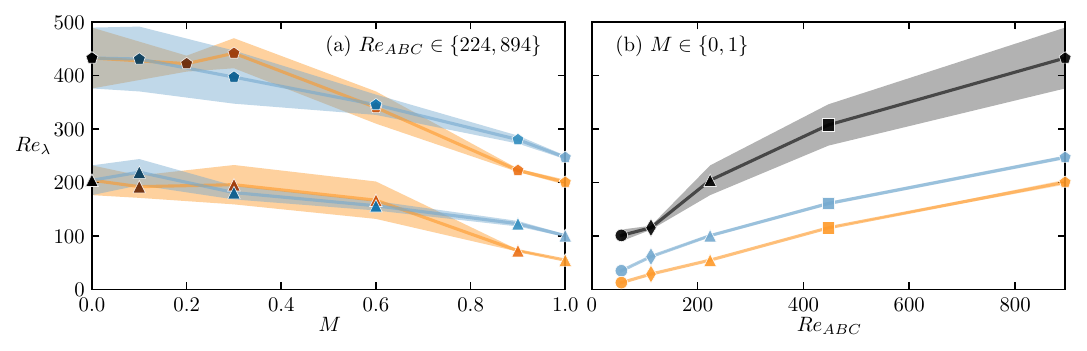}
\caption{(a) Effect of the solid mass fraction $M$ on the Taylor Reynolds number $Re_\lambda$. Flows with $Re_{ABC}=224$ are marked using triangles, and flows with $Re_{ABC}=894$ are marked using pentagons. (b) Effect of the forcing Reynolds number $Re_{ABC}$ on the Taylor Reynolds number $Re_\lambda$ for the single phase and particle-laden cases with fixed particles ($M=1$). In both plots, we show the single-phase flows in black, flows with spheres in blue, and flows with fibres in yellow; the shaded regions give the root-mean-square of $Re_\lambda$ in time. {Each case is plotted using its marker, which is listed in table~\ref{tabSinglePhCases}, \ref{tabSphCases}, or \ref{tabFibCases}.}}
 \label{fig:Re}
\end{figure}

The Taylor Reynolds number is defined as {$Re_\lambda\equiv u_{rms} \lambda / \nu $, where $u_{rms}\equiv\sqrt{\langle u_i u_i \rangle_{\mathbf x,t}}/3$} is the root-mean-square velocity, $\lambda=u_{{rms}}\sqrt{15\nu/\epsilon}$ is the Taylor length scale, {$\epsilon\equiv 2\nu \langle s_{ij} s_{ij}\rangle_{\mathbf{x},t}$} is the viscous dissipation rate, {${s_{ij}\equiv (\partial_i u_j + \partial_j u_i)/2}$ is the strain-rate tensor, and angled brackets $\langle\cdot\rangle_{\mathbf{x},t}$ indicate an average over space $\mathbf{x}$ and time $t$.}

The Taylor-Reynolds number is an indicator of the intensity of turbulence in the flow, and in figure~\ref{fig:Re}, we show how $Re_\lambda$ compares for all of our cases. Figure~\ref{fig:Re}a shows that increasing the solid mass fraction $M$ causes a reduction in $Re_\lambda$ at both high and low forcing Reynolds numbers. Both spheres and fibres produce a comparable decrease in $Re_\lambda$, but the reduction effect is more substantial for very heavy fibres. Looking at the trend in $Re_\lambda$ with the forcing Reynolds number $Re_{ABC}$ in figure~\ref{fig:Re}b, we see, as expected, a monotonic increase in all three cases, i.e., single phase, sphere-laden, and fibre-laden flows.

\begin{figure}
\centering
\includegraphics[width=\textwidth]{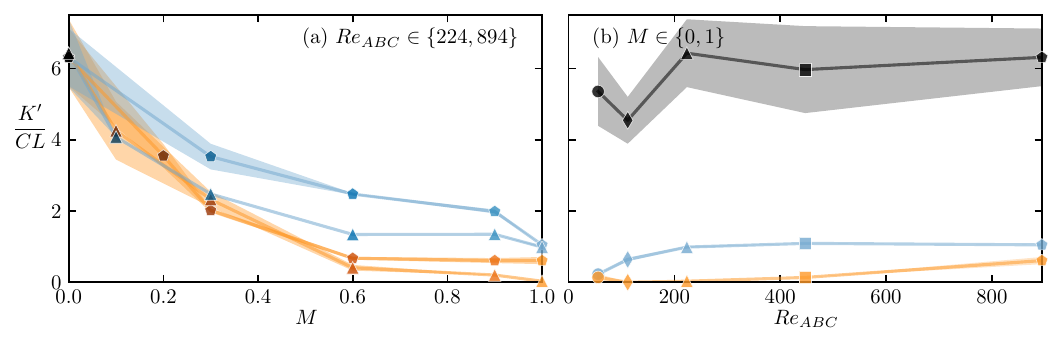}
\caption{{The turbulent kinetic energy $K'$, normalised using the forcing amplitude $C$ and length $L$. (a) shows the effect of solid mass fraction $M$, and (b) shows the effect of forcing Reynolds number $Re_{ABC}$. Each case is plotted using its marker, which is listed in table~\ref{tabSinglePhCases}, \ref{tabSphCases}, or \ref{tabFibCases}. We show the single-phase flows in black, flows with spheres in blue, and flows with fibres in yellow; the shaded regions give the root-mean-square of $K'$ in time.}}
\label{fig:flucEVsM}
\end{figure}
{A second way to quantify turbulence is the turbulent kinetic energy $K' \equiv \langle u_i' u_i' \rangle_{\mathbf{x},t}/2$. Similarly to \textcite{oka_attenuation_2022-1}, we compute $K'$ using the fluctuating part of the fluid velocity $u_i'(\mathbf{x},t)\equiv u_i(\mathbf{x},t)-\langle{ u_i(\mathbf{x,t}) \rangle_t}$. Figure~\ref{fig:flucEVsM}a shows the dependence of turbulent kinetic energy on particle mass fraction. We see that adding spheres and fibres reduces $K'$ relative to the single-phase flow, with fibres having a slightly greater attenuation effect at both high and low forcing Reynolds numbers. From figure~\ref{fig:flucEVsM}b, we see that the turbulent kinetic energy shows only a weak dependence on $Re_{ABC}$, the single-phase $(M=0)$ cases remain roughly constant (around $K'\approx 6CL$), and flows with fixed (M=1) particles show a large attenuation of $K'$ for all $Re_{ABC}$. This agrees with the observations of \textcite{peng_parameterization_2023}, who made simulations of spherical particles in homogeneous isotropic turbulence and found that the attenuation of turbulent kinetic energy has little dependence on the Reynolds number.}

\begin{figure}
\centering
\includegraphics[width=\textwidth]{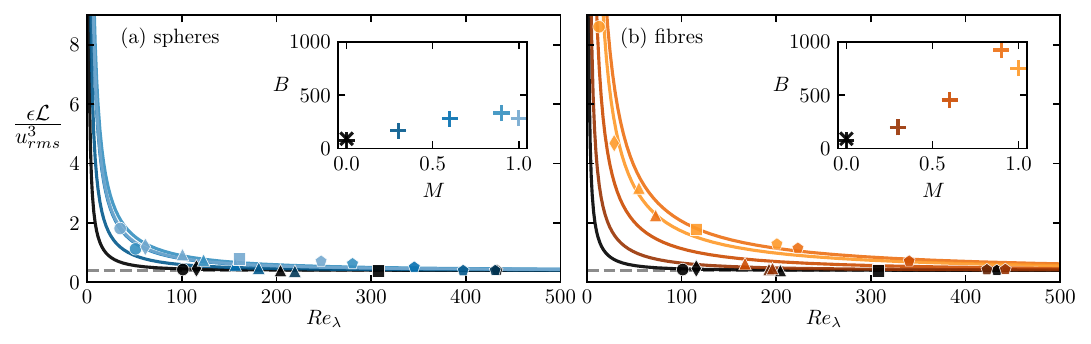}
\caption{The dependence of the normalised dissipation $\epsilon \mathcal{L}/u_{{rms}}^3$ on the Taylor-Reynolds number $Re_\lambda$. Flows with spheres are marked blue on panel (a), flows with fibres are marked in orange on panel (b), and single-phase flows are marked in black on both panels. {Each case is plotted using its marker, which is listed in table~\ref{tabSinglePhCases}, \ref{tabSphCases}, or \ref{tabFibCases}.} Dashed lines show the anomalous value of dissipation $\epsilon \mathcal{L}/u_{{rms}}^3=0.4$ measured by \citet{donzis_scalar_2005}. Solid lines show fits of equation~\ref{eq:donz} with $A=0.2$. The fitted values of $B$ are given in the inset, where Donzis' result $(B=92)$ is marked with an ``X''.}
 \label{fig:donz}
\end{figure}
The dissipative anomaly is a well-studied feature in single-phase turbulent flows, first proposed by~\citet{taylor_statistical_1935}. It states that, even in the limit of vanishing viscosity ($Re_\lambda\to\infty$), the normalised dissipation $\epsilon \mathcal{L}/u_{{rms}}^3$ remains finite. The integral length scale is given by,
\begin{equation}
\mathcal{L}=\frac{\pi}{2u_{{rms}}^2}\int_0^{\infty}\frac{E}{\kappa}\mathrm{d}\kappa,
\end{equation}
where $E$ is the turbulent kinetic energy spectrum and $\kappa$ is the wavenumber. {In 2005,} \citet{donzis_scalar_2005} parametrized the dissipative anomaly, based on the {mathematical derivation} of \citet{doering_energy_2002}, they fit the function 
\begin{equation}
\frac{\epsilon \mathcal{L}}{u_{{rms}}^3}=A\left[1+\sqrt{1+(B/Re_\lambda)^2}\right]
\label{eq:donz}
\end{equation}
to a number of single-phase flows, obtaining $A=0.2$ and $B=92$. Figure~\ref{fig:donz} shows the dependence of the normalised dissipation for our flows. We see that as $Re_\lambda \to \infty$, flows with spheres and fibres at all mass fractions appear to converge to the same anomalous value of dissipation $\epsilon \mathcal{L}/u_{{rms}}^3{\to}0.4$. Hence, we fit equation~\ref{eq:donz} with $A=0.2$ to each mass fraction of spheres and fibres, obtaining a value for $B$ in each case, shown in the insets of figure~\ref{fig:donz}. The fit to our single phase flows agrees closely with~\citet{donzis_scalar_2005}'s result. However, both spheres and fibres cause an increase in the value of $B$ as their mass fraction increases, with fibres producing roughly double the effect. To understand how spheres and fibres modify the dissipation in the flow, we must look at how they transport energy from large to small scales into the dissipative range.

\subsection{Scale-by-scale results}
\begin{figure}
	\centering 
	\includegraphics[width=\textwidth]{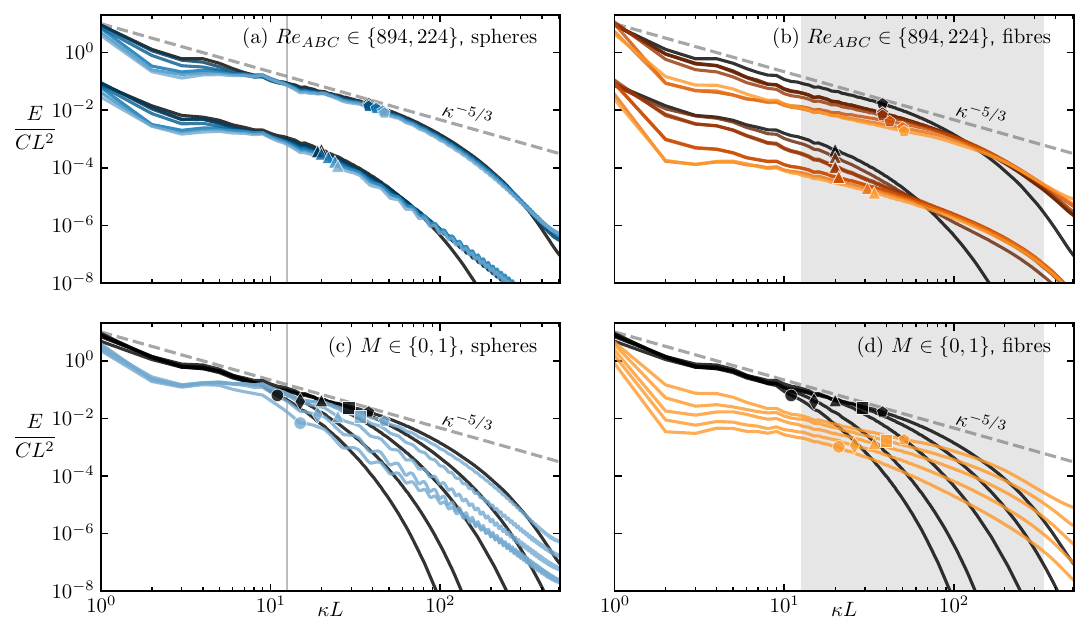}
	\caption{Energy spectra of all cases, {normalised using the forcing amplitude $C$ and length $L$.} On the left, we show flows with spheres, and the vertical grey lines show the wavenumber $\kappa_c=2\pi/c$ of the sphere diameter. On the right, we show flows with fibres, and we shade the region between the wavenumber $\kappa_c$ of the fibre length and the wavenumber $\kappa_d=2\pi/d$ of the fibre diameter. (a) and (b) show $Re_{ABC}=894$ and $Re_{ABC}=224$, with the latter shifted downwards by a factor of 100 on the y-axis, for various mass fractions. (c) and (d) show fixed $(M=1)$ and single phase $(M=0)$ cases for various Reynolds numbers. {Each case is marked at the wavenumber corresponding to the Taylor length scale, using the marker listed in table~\ref{tabSinglePhCases}, \ref{tabSphCases}, or \ref{tabFibCases}.}}
	\label{fig:spectra}
\end{figure}

Figure~\ref{fig:spectra} shows each flow's turbulent kinetic energy spectrum $E$. Single-phase flows exhibit the canonical Kolmogorov scaling $E\sim \kappa^{-5/3}$ for one or two decades, depending on the forcing Reynolds number. Adding spheres reduces $E$ at wavenumbers up to the sphere diameter ($\kappa<\kappa_c$) and increases $E$ in the dissipative range. We mention, in passing, the oscillations at the wavenumber of the sphere diameter; these are an artefact resulting from the discontinuity in the velocity gradient at the sphere boundary~\citep{lucci_modulation_2010}.
The addition of fibres causes a reduction in $E$ across a broad band of wavelengths ($\kappa L\lesssim100$) and a pronounced increase in the dissipative range. As was previously seen by~\citet{olivieri_effect_2022}, the energy scaling $E\sim \kappa^{-\beta}$ becomes flatter as the mass fraction $M$ of fibres increases. Figure~\ref{fig:spectra} also shows that both spheres and fibres cause the Taylor length scale to shift to higher wavenumbers due to the increase of the energy in the dissipative scales due to the presence of particles.

\begin{figure}
\centering
\includegraphics[width=\textwidth]{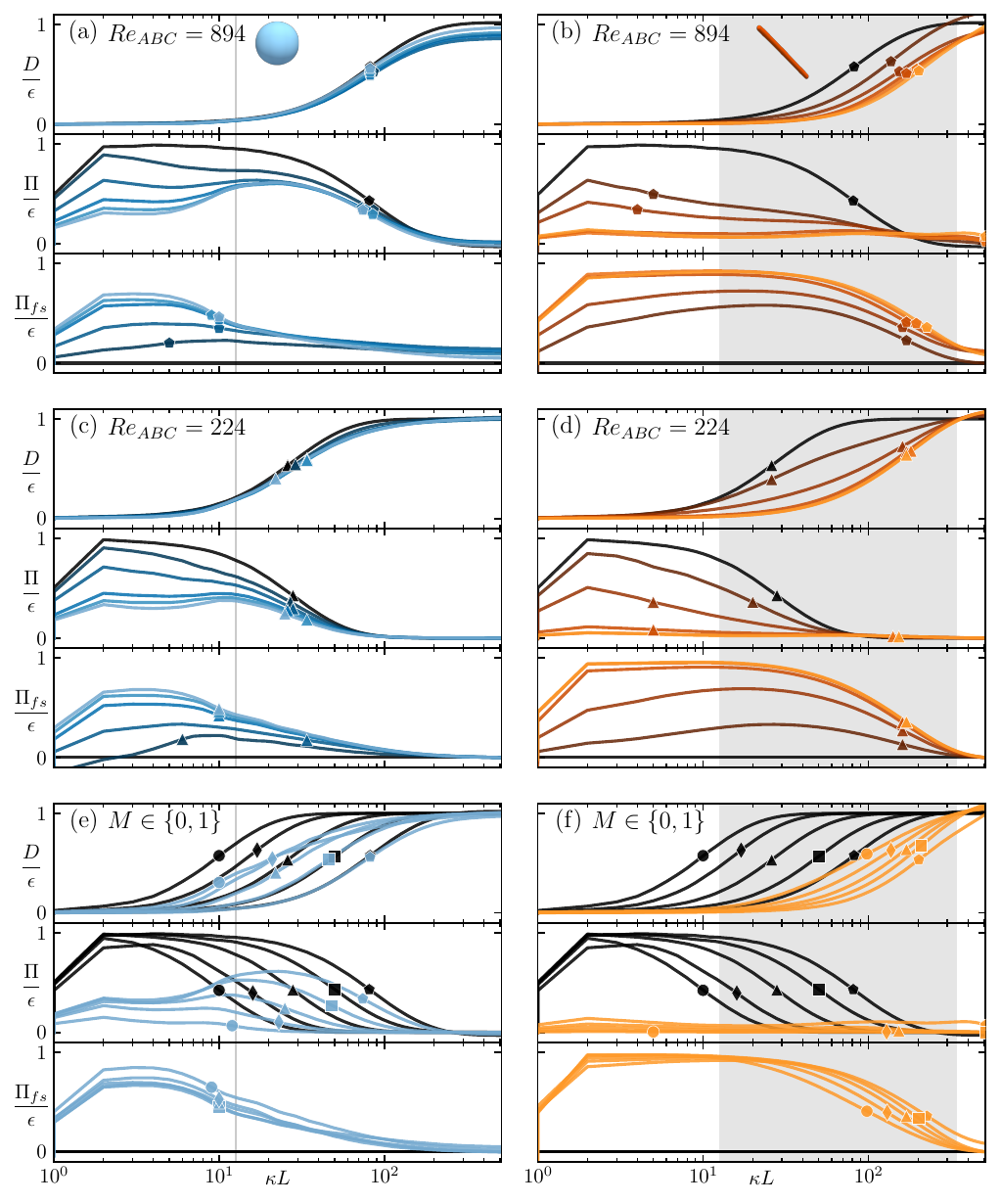}
\caption{Scale-by-scale energy balance for all cases. Panels (a) and (b) show $Re_{ABC}=894$ cases for various mass fractions. Panels (c) and (d) show $Re_{ABC}=224$ cases for various mass fractions. Panels (e) and (f) show fixed ($M=1$) and single-phase ($M=0$) cases for various Reynolds numbers. On the left, we show flows with spheres, and the vertical grey lines show the wavenumber $\kappa_c=2\pi/c$ of the sphere diameter. On the right, we show flows with fibres, and we shade the region between the wavenumber $\kappa_c$ of the fibre length and the wavenumber $\kappa_d$ of the fibre diameter. For each case, we plot three terms: the Dissipation $D$, the energy flux $\Pi$ due to convection, and the energy flux $\Pi_{sf}$ due to the solid-fluid coupling. Each curve is marked where the gradient is largest, {using the marker listed in table~\ref{tabSinglePhCases}, \ref{tabSphCases}, or \ref{tabFibCases}.}}
 \label{fig:flux}
\end{figure}

Figure~\ref{fig:flux} shows how each term in the Navier-Stokes equation interacts with the energy spectrum, {as} expressed by the spectral energy balance,
\begin{equation}
\mathcal{F}_{inj}(\kappa)+\Pi(\kappa)+\Pi_{sf}(\kappa)+{D}(\kappa)=\epsilon,
\end{equation}
where $\mathcal{F}_{inj}, \Pi, \Pi_{sf},$ and $D$ are the rate of energy transfer by the ABC forcing, advection, solid-fluid coupling, and dissipation, respectively. See the supplementary information of Ref.~\cite{abdelgawad_scaling_2023} for a derivation of this equation. For the single-phase flows, energy is carried by the advective term $\Pi$ from large to small scales, where it is dissipated by the viscous term $D$. When particles are added, we see that the solid-fluid coupling term $\Pi_{sf}$ acts as a ``spectral shortcut''~\cite{finnigan_turbulence_2000,olivieri_effect_2022}; it bypasses the classical turbulent cascade, removing energy from large scales and injecting it at the length scale of the particles, through their wakes. In keeping with the spectra, the power of the solid-fluid coupling (shown by the peak value of $\Pi_{sf}$) increases with mass fraction $M$ of both spheres and fibres. For spheres, the coupling $\Pi_{sf}$ dominates only at wavenumbers less than that of the sphere diameter $\kappa<\kappa_c$ (i.e., large length scales). Around $\kappa_c$, the sphere wake returns energy to the fluid, and the classical cascade resumes for $\kappa>\kappa_c$. For fibres instead, $\Pi_{sf}$ extends deep into the viscous range. In fact, markers on the $\Pi_{sf}$ curves show that spheres mostly inject energy around the length scale of their diameter, and fibres mostly inject energy at a length scale between their thickness and length. The images of dissipation in figure~\ref{fig:snapshots} {support this observation}; around the spheres, we see wakes comparable in size to the {sphere} diameter, while around the fibres, we see wakes comparable in size to the {fibre} thickness. We presume that the fluid-sphere coupling $\Pi_{sf}$ would extend into the viscous range if the sphere diameter was {smaller}. Lastly, we consider the effect of Reynolds number on energy transfers in the flow, reducing $Re_{ABC}$ limits the range of scales at which the advective flux $\Pi$ occurs for single-phase flows. However, $Re_{ABC}$ has little effect on the wavenumber range of the solid-fluid coupling term $\Pi_{sf}$, suggesting that the size of the particle wakes is governed mainly by particle geometry, with a lesser effect from fluid properties like viscosity.

\begin{figure}
\centering
\includegraphics[width=\textwidth]{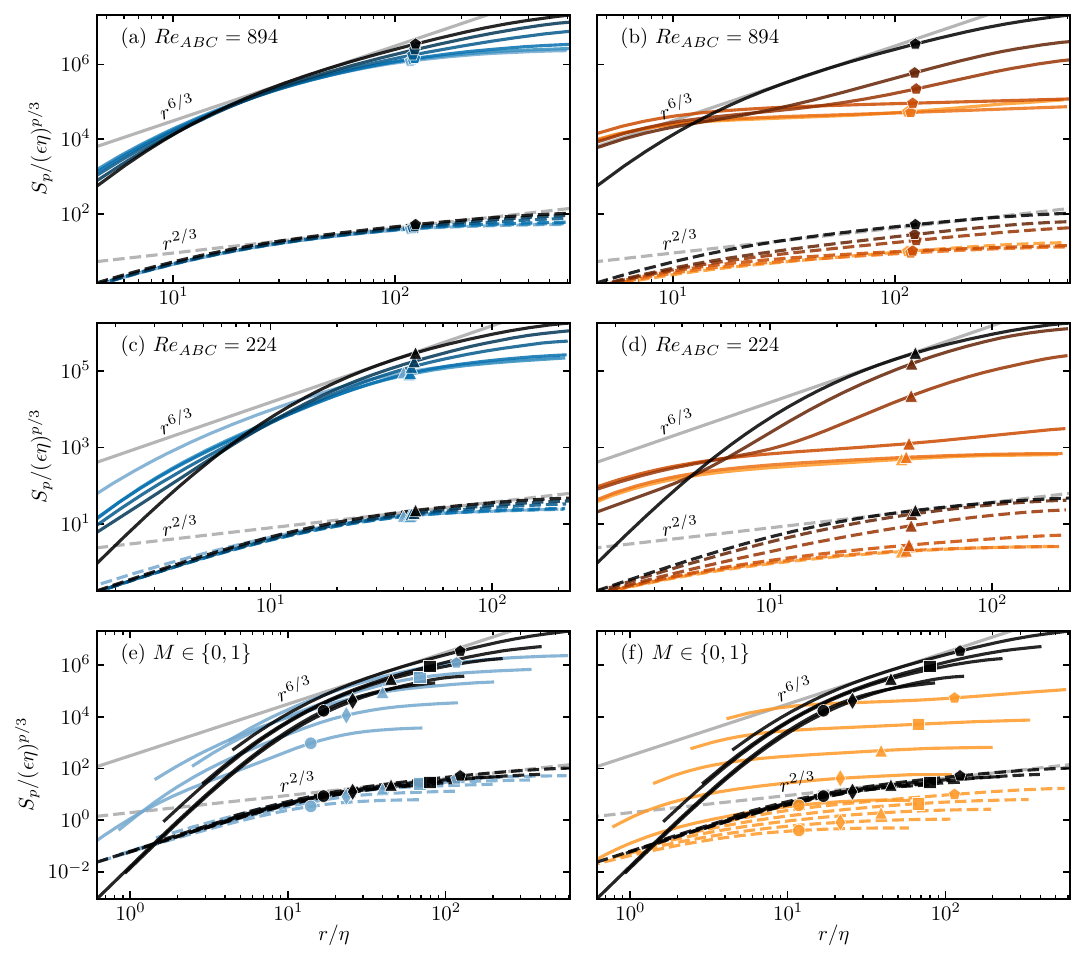}
\caption{Structure functions of order $p=2$ (dashed lines) and $p=6$ (solid lines). Panels (a) and (b) show cases with $Re_{ABC}=894$ for various mass fractions. Panels (c) and (d) show cases with $Re_{ABC}=224$ for various mass fractions. Panels (e) and (f) show fixed ($M=1$) and single-phase ($M=0$) cases for various Reynolds numbers. Flows with spheres are on the left, and flows with fibres are on the right. Grey lines show the scalings predicted by \citet{kolmogorov_local_1941} for the single-phase structure functions. We mark each line at $r=c$ {using the marker listed in table~\ref{tabSinglePhCases}, \ref{tabSphCases}, or \ref{tabFibCases}.}}
 \label{fig:strFun}
\end{figure}

Moving from wavenumber space to physical space, we show the longitudinal structure functions 
\begin{equation}
S_p{(r)}\equiv \langle \left[ \hat r_i u_i(\mathbf x+\mathbf r{,t})- \hat r_i u_i(\mathbf{x}{,t}) \right]^p\rangle_{{\mathbf x,\mathbf{\hat r},t}}
\end{equation}
of each flow in figure~\ref{fig:strFun}, where $\mathbf r$ is a separation vector between two points in the flow, it has magnitude $r$ and direction $\mathbf{\hat r}$, and angled brackets show an average over space $\mathbf{x}$, {direction $\mathbf{\hat r}$, and time t}. For the second moment $(p=2)$, the single-phase flows closely follow Kolmogorov's scaling~\cite{kolmogorov_local_1941} in the inertial range ($r\gg\eta$). However, when particles are added, $S_2$ decreases relative to the single-phase case. For spheres, the decrease occurs for $r\gtrsim c$, while for fibres, it occurs at even smaller separations $r$. Much like the $E\sim k^{-\beta}$ scalings seen in figure~\ref{fig:spectra}, the decreased regions in figure~\ref{fig:strFun} show scalings $S_2\sim r^{\zeta_2}$, which become flatter for larger mass fractions $M$. These energy spectrum and structure-function scalings are, in fact, just observations of the same phenomenon, as the exponents are related by $\zeta_2=\beta-1$ for any differentiable velocity field \citep[p.232]{pope_turbulent_2000}. 
In the higher moment structure function ($p=6$), the flattening of the $S_6\sim r^{\zeta_6}$ scaling by the particles is yet more apparent, indicating that the tails of the probability distribution of $ \hat r_i u_i(\mathbf x+\mathbf r)- \hat r_i u_i(\mathbf{x}) $ become wider for smaller separations $r$. In other words, extreme values become more common, and the velocity field becomes more intermittent in space. In the case of single-phase homogeneous-isotropic turbulence, the intermittency of the velocity field has been {linked} to the non-space-filling nature of the fluid dissipation \citep{frisch_turbulence_1995}. Next, we explore this link in the case of particle-laden turbulence by investigating the intermittency of dissipation $\epsilon$ in space. 

\begin{figure}
\centering
\includegraphics[width=\textwidth]{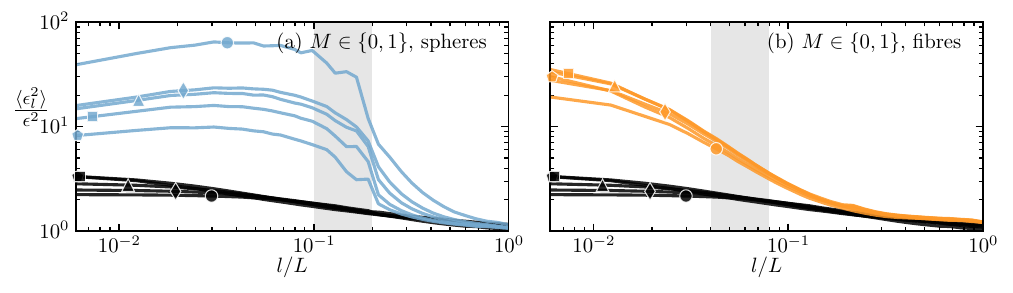}
\caption{{The fluid dissipation averaged in balls of radius $l$ and squared. We mark each case at $l=\eta$ using the marker listed in table~\ref{tabSinglePhCases}, \ref{tabSphCases}, or \ref{tabFibCases}. We shade the regions where $\langle\epsilon_l^2\rangle$ shows a strong dependence on $l$, for sphere-laden flows (a) this is $0.1L\leq l\leq0.2L$, whereas for fibre-laden flows (b) this is $0.04L\leq l\leq0.08L$.}}
 \label{fig:dissMom}
\end{figure}

{Using the method described by~\citet[p.159]{frisch_turbulence_1995}, \citet{meneveau_multifractal_1991-1} and others, we obtain the quantity $\epsilon_l$ by averaging the viscous dissipation within a spherical region of radius $l$ (hereafter, we refer to this averaging region as a ``ball'' to avoid confusion with the particles). To observe the variation of $\epsilon_l$ we take its square and average over many balls of radius $l$ at different locations and times, where the average is denoted using angled brackets $\langle\cdot\rangle$. Figure~\ref{fig:dissMom} shows how $\langle\epsilon_l^2\rangle$ depends on the radius $l$ of the balls. At large radii $(l\to\infty)$, $\epsilon_l$ is by definition equivalent to the bulk value $\epsilon$, and so $\langle\epsilon_l^2\rangle/\epsilon^2$ tends to unity. However, for $l<L$, we see that $\langle\epsilon_l^2\rangle/\epsilon^2>1$ in all of the cases plotted. This increase indicates that there are localised regions of high dissipation in the fluid. In other words, the dissipation fields are intermittent in space. For the single-phase flows, $\langle\epsilon_l^2\rangle$ reaches a plateau for $l<\eta$, giving an indication of the size of these flow structures. For the sphere-laden flows, we see that $\langle\epsilon_l^2\rangle$ reaches almost 100 times the bulk value and remains roughly constant for small radii $(l<0.05L)$. This suggests that the spherical particles are creating high dissipation structures in the fluid, and balls with $l<0.05L$ fit entirely inside these structures. That is, reducing the ball radius below $l=0.05L$ has minimal effect on $\langle\epsilon_l^2\rangle$ because the majority of balls are sampling either entirely inside a high dissipation structure or entirely outside. In the case of fibres, high dissipation structures also produce an increase in $\langle\epsilon_l^2\rangle$ for small $l$; here no plateau is seen, implying that even the smallest balls $(l=6\times10^{-3}L)$ cannot fit inside the high dissipation structures created by the fibres.}

\begin{figure}
\centering
\includegraphics[width=\textwidth]{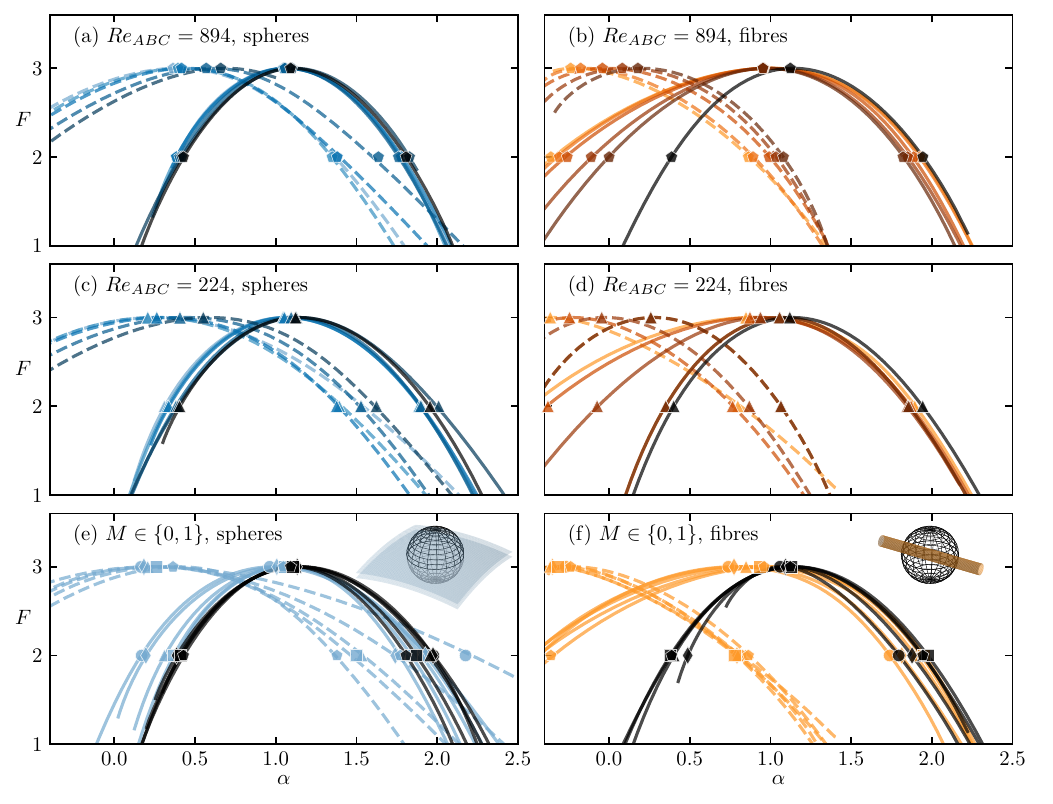}
\caption{Multifractal spectra of the dissipation in regions containing particles (dashed lines) and regions not containing particles (solid lines). {Spectra in the left panels (a,c,e) were calculated using balls of radius $0.1L\leq l\leq 0.2L$. Spectra in the right panels (b,d,f) were calculated using balls of radius $0.04L\leq l\leq 0.08L$. We mark each line case where $F$ is maximum and at $F=2$ using the marker listed in table~\ref{tabSinglePhCases}, \ref{tabSphCases}, or \ref{tabFibCases}.} Panels (a) and (b) show cases with $Re_{ABC}=894$ for various mass fractions. Panels (c) and (d) show cases with $Re_{ABC}=224$ for various mass fractions. Panels (e) and (f) show fixed ($M=1$) and single-phase ($M=0$) cases for various Reynolds numbers. {Panel (e) shows a two-dimensional sheet intersecting with a ball, and panel (f) shows a one-dimensional tube intersecting with a ball.}}
\label{fig:fractality}
\end{figure}

{To measure the fractal dimension of the high dissipation structures, we take a range of moments $-6\le q\le6$ of $\epsilon_l$. This quantity has a scaling behaviour 
\begin{equation}
\langle\epsilon_l^q\rangle\sim l^{\tau},
\label{eqDissScaling}
\end{equation}
if the dissipation field is multifractal~\cite{frisch_turbulence_1995}. Indeed, for the sphere-laden flows in figure~\ref{fig:dissMom}a, $\langle\epsilon_l^2\rangle$ has a strong dependence on $l$ in the range $0.1L\le l\le0.2L$, and for the fibre-laden flows in figure~\ref{fig:dissMom}b, $\langle\epsilon_l^2\rangle$ has a strong dependence on $l$ in the range $0.04L\le l\le0.08L$. Hence, we calculate $\tau$ by making lines of best fit in these ranges, and we obtain the multifractal spectra $F(\alpha)$ with a Legendre transformation
\begin{align}
\alpha&=\frac{d\tau}{d q}+1,\label{eqLegendreA}\\
F&=q(\alpha-1)-\tau+3.\label{eqLegendreF}
\end{align}
}
Figure~\ref{fig:fractality} shows the multifractal spectra for all cases. 
Similarly to~\citet{mukherjee_turbulent_2023}, we split our analysis into separate regions: solid curves result from an ensemble average over balls not containing particles, while dashed curves result from an ensemble average over balls containing particles. We see that the multifractal spectra of the single-phase flows and the regions not containing particles have peaks at $\alpha\approx1, F\approx3$, showing there is a background of space-filling dissipation in these regions~\citep[p.163]{frisch_turbulence_1995}. {In addition, the presence of tails in the spectra shows that the dissipation fields are not self-similar and violate the hypotheses made by~\citet{kolmogorov_local_1941} for a turbulent flow. This can explain the flattening of the high-order structure functions ($S_6$ in figure~\ref{fig:strFun}) relative to Kolmogorov's predictions. We note in passing that the single-phase flows with lower Reynolds number ($Re_{ABC}=55.9$ and $Re_{ABC}=112$) show narrower tails in figure~\ref{fig:fractality}f, as viscous effects are present here.}

The multifractal spectra of the regions containing spheres (blue dashed lines in figure~\ref{fig:fractality}) have peaks which are shifted to the left. This shift can be explained by considering the high dissipation in the boundary layers around the spheres. As imaged in figure~\ref{fig:dissSurf}, the boundary layers {have a higher dissipation rate than the surrounding fluid, and} are confined in space to relatively thin sheets. {In figure~\ref{fig:fractality}e we sketch a ball of radius $l$ intersecting with a sphere's boundary layer which has thickness $\delta\ll l$. The volume of the high dissipation region inside the ball is $\pi l^2\delta$, hence the total dissipation in the ball roughly} scales as $\sim l^2$. When we average over the volume of the ball (${4\pi l^3/3}$) and take the $q$th moment, we obtain $\langle\epsilon_l^q\rangle\sim l^{2q}l^{-3q}$. {Comparing with equation~\ref{eqDissScaling}, this scales as $\tau=-q$. Applying a Legendre transformation (equations~\ref{eqLegendreA} and~\ref{eqLegendreF}) this corresponds to} the point $\alpha=0, F=3$, which is roughly where {we find the peaks in the multifractal spectra of the regions containing spheres. This reinforces our observation that the dissipation in the regions around the spherical particles is confined to sheet-like structures.}

The multifractal spectra of the regions containing fibres (orange dashed lines in figure~\ref{fig:fractality}) have peaks which are shifted even further to the left. Again, we explain this by considering the {boundary layers} around the particles. {In figure~\ref{fig:fractality}f we sketch a ball of radius $l$ intersecting with a fibre's boundary layer which has radius $\delta\ll l$}. In this case, the volume of {the high dissipation region is $\pi\delta^2 l$. Thus, the average dissipation in the ball scales as} $\langle\epsilon_l^q\rangle\sim l^{q}l^{-3q}$, so the resulting scaling is $\tau=-2q$, which (by {equations~\ref{eqLegendreA} and~\ref{eqLegendreF}) corresponds to} the point $\alpha=-1, F=3$. In fact, {in figure~\ref{fig:fractality}f}, we see that the ensemble average over balls containing $M=1$ fibres produces peaks {around $\alpha=-0.5$}, suggesting that the wakes around the fibres are not exactly thin tubes, but more space-filling structures with a {fractal dimension between one and two}. This is also supported by the images in figure~\ref{fig:dissSurf}, as wakes are seen to extend behind the fibres.

The middle and upper panels of figure~\ref{fig:fractality} show that (at $Re_{ABC}=224$ and $Re_{ABC}=894$) the sphere- and fibre-containing multifractal spectra tend toward the single phase multifractal spectrum as $M$ is reduced. This is expected as the particle-fluid relative velocity is reduced and the wakes become less prominent compared to the background space-filling turbulence. On the other hand, the ensembles of balls not containing particles produce multifractal spectra with low $\alpha$ tails that extend beyond those of the single-phase multifractal spectra, suggesting that the wakes of spheres and fibres extend into the bulk of the fluid and that they retain their non-space-filling nature.

\subsection{Local flow structures}
\begin{figure}
\centering
\includegraphics[width=\textwidth]{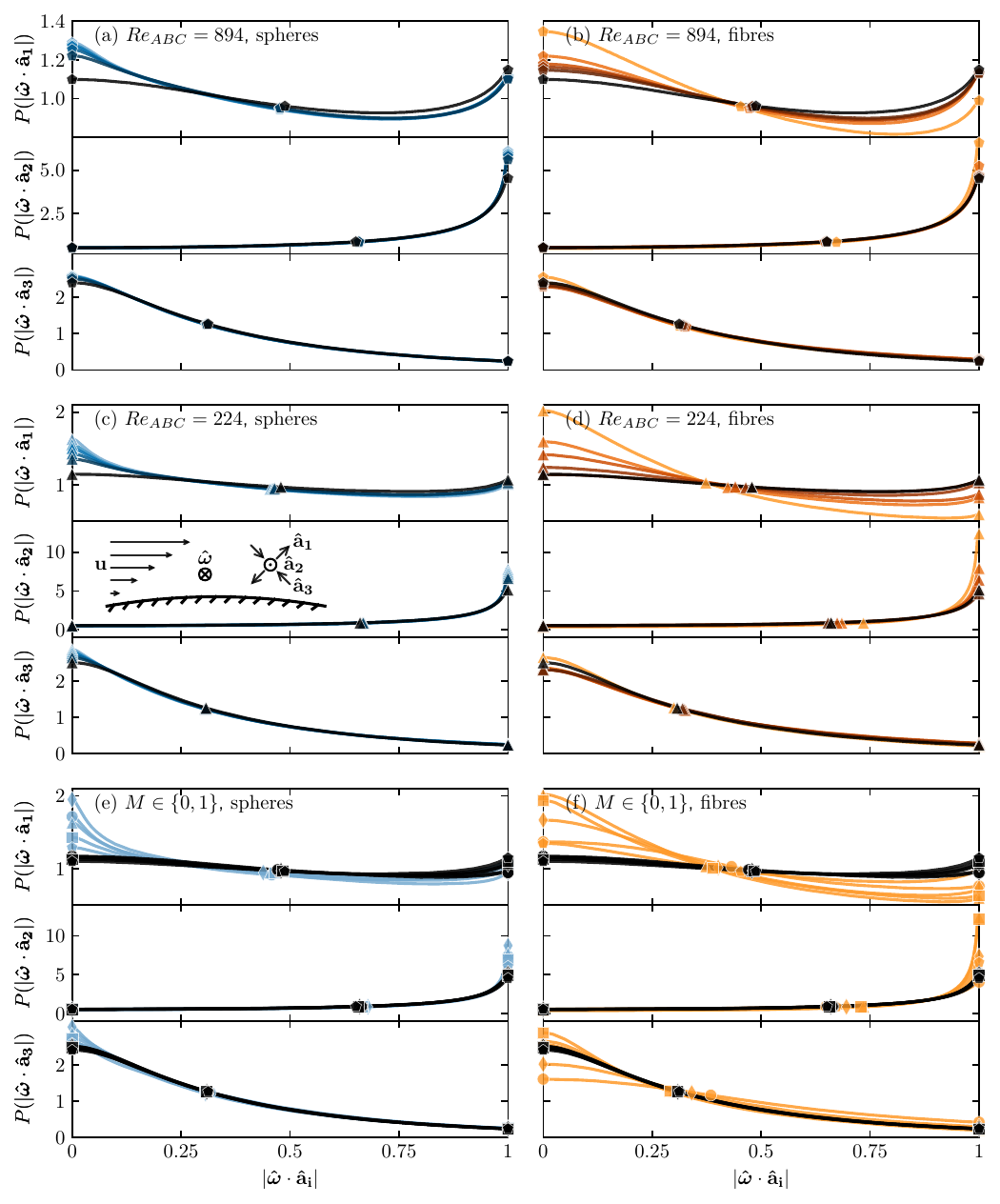}
\caption{Histograms of the alignment of the vorticity unit vector $\boldsymbol{\hat{\omega}}$ with the eigenvectors $\mathbf{\hat{a}_1},\mathbf{\hat{a}_2},\mathbf{\hat{a}_3}$ of the strain-rate tensor. Flows with spheres are on the left, and flows with fibres are on the right. Panels (a) and (b) show cases with $Re_{ABC}=894$ for various mass fractions. Panels (c) and (d) show cases with $Re_{ABC}=224$ for various mass fractions. Panels (e) and (f) show fixed ($M=1$) and single-phase ($M=0$) cases for various Reynolds numbers. {Each case is plotted using its marker, which is listed in table~\ref{tabSinglePhCases}, \ref{tabSphCases}, or \ref{tabFibCases}.} We mark each line at $|\boldsymbol{\hat\omega}\cdot\mathbf{\hat a_i}|=0,1$ and $\langle|\boldsymbol{\hat\omega}\cdot\mathbf{\hat a_i}|\rangle$. The inset of panel ({c}) shows the directions of vorticity and the eigenvectors of the strain rate in a laminar shear flow, where $\mathbf{\hat{a}_1}$ and $\mathbf{\hat{a}_3}$ are confined to the shear plane, while $\mathbf{\hat{a}_2}$ and $\boldsymbol{\hat{\omega}}$ are perpendicular to it.}
 \label{fig:align}
\end{figure}

To look more closely at the flow structures produced by the particles, we compute the alignment of the unit-length eigenvectors $\mathbf{\hat a_1,\hat a_2,\hat a_3}$ of the strain-rate tensor ${s_{ij}\equiv (\partial_i u_j + \partial_j u_i)/2}$ with the vorticity ${\boldsymbol\omega\equiv{\varepsilon_{ijk}}\partial_j u_k\mathbf{\hat{e}}_i}$~\citep{ashurst_alignment_1987,olivieri_fully_2022}, where $\mathbf{\hat e_1,\hat e_2,\hat e_3}$ are the Cartesian unit vectors, {and $\varepsilon_{ijk}$ is the Levi-Civita symbol}. We choose $\mathbf{\hat a_1}$ to be the eigenvector corresponding to the largest eigenvalue. For an incompressible fluid, the three eigenvalues of $s_{ij}$ sum to zero, so the largest eigenvalue is never negative. Hence, $\mathbf{\hat a_1}$ aligns with the direction of elongation in the flow. Similarly, we choose the eigenvector $\mathbf{\hat a_3}$ to correspond with the smallest eigenvalue, which is never positive, so $\mathbf{\hat a_3}$ aligns with the direction of compression in the flow. Finally, due to the symmetry of the strain-rate tensor, $\mathbf{\hat a_2}$ is orthogonal to the other two eigenvectors and, depending on the flow, there can be compression or elongation along its axis. Figure~\ref{fig:align} shows probability-density functions $P$ of the scalar product of the unit-length vorticity $\boldsymbol{\hat\omega}$ with $\mathbf{\hat a_1,\hat a_2}$ and $\mathbf{\hat a_3}$ for all of the flows studied. The quantity $\boldsymbol{\hat\omega}\cdot\mathbf{\hat a_i}$ is simply the cosine of the angle between the two vectors, and we take the modulus because $\mathbf{\hat a_i}$ and $-\mathbf{\hat a_i}$ are degenerate eigenvectors. In the single-phase flows, $P$ is maximum at $|\boldsymbol{\hat\omega}\cdot\mathbf{\hat a_2|}=1$, that is, we see a strong alignment of vorticity with the intermediate eigenvector $\mathbf{\hat a_2}$. This is a well-known feature of turbulent flows {and} has been attributed to the axial stretching of vortices~\citep{ashurst_alignment_1987}. Also in keeping with previous observations of single-phase turbulence, we see that the first eigenvector $\mathbf{\hat a_1}$ shows very little correlation with $\boldsymbol{\hat\omega}$, and the last eigenvector is mostly perpendicular to the vorticity, producing a peak in $P$ at $\boldsymbol{\hat\omega}\cdot\mathbf{\hat a_3}=0$. When spheres are added, vorticity aligns even more strongly with the intermediate eigenvector $\mathbf{\hat a_2}$, and becomes more perpendicular to the first and last eigenvectors $\mathbf{\hat a_1}$ and $\mathbf{\hat a_3}$. This change is consistent with the existence of a shear layer at the surface of the spheres, {or in the wakes behind the spheres}. To demonstrate this, we draw a laminar shearing flow and label the directions of $\mathbf{\hat a_1,\hat a_2,\hat a_3}$ and $\boldsymbol{\hat\omega}$ in the inset of figure~\ref{fig:align}. The compression $\mathbf{\hat a_3}$ and extension $\mathbf{\hat a_1}$ are in the plane of the shear flow, while the vorticity $\boldsymbol{\hat\omega}$ and the intermediate eigenvector $\mathbf{\hat a_3}$ are perpendicular to the shear plane; this gives rise to the $\boldsymbol{\hat\omega}\cdot\mathbf{\hat a_1}=0,$ $|\boldsymbol{\hat\omega}\cdot\mathbf{\hat a_2}|=1$ and $\boldsymbol{\hat\omega}\cdot\mathbf{\hat a_3}=0$ modes in the sphere-laden flows in figure~\ref{fig:align}. The larger mass fraction spheres have a greater effect, as their motion relative to the fluid is larger. Also, the spheres' effect is more pronounced at lower $Re_{ABC}$, which can be explained by an increased thickness in the shear layer as the fluid {viscosity increases}. For fibre-laden flows, peaks are also seen at  $\boldsymbol{\hat\omega}\cdot\mathbf{\hat a_1}=0,$ $|\boldsymbol{\hat\omega}\cdot\mathbf{\hat a_2}|=1$ and $\boldsymbol{\hat\omega}\cdot\mathbf{\hat a_3}=0$, but the peaks are spread over a larger range of angles, presumably due to the high curvature of the fibre surfaces, which adds variance to the local direction of vorticity and shear. At small Reynolds numbers ($Re_{ABC}=55.9$ and $112$) the peaks are widest, and $P(|\boldsymbol{\hat\omega}\cdot\mathbf{\hat a_i}|)$ becomes an almost uniform distribution.

\begin{figure}
\centering
\includegraphics[width=\textwidth]{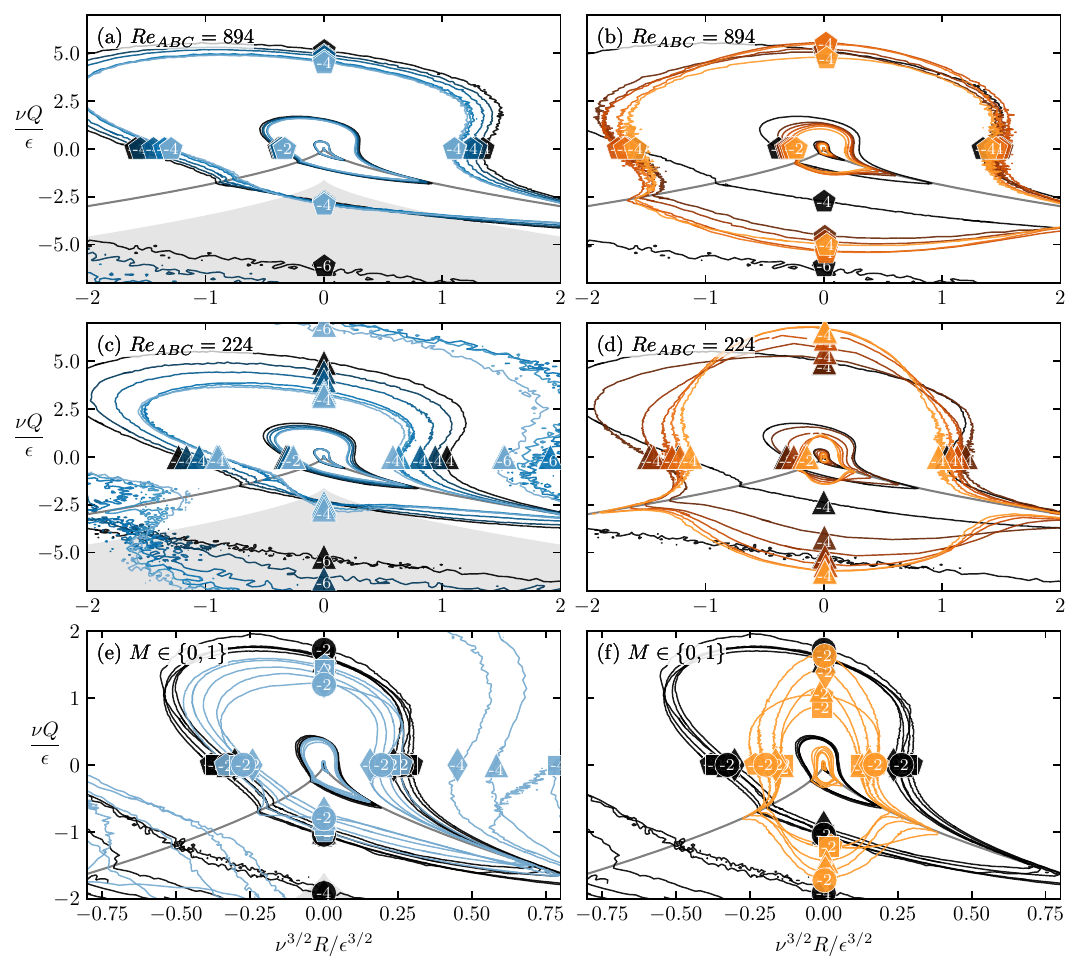}
\caption{Joint histograms of the invariants $Q$ and $R$ of the fluid velocity gradient tensor. Panels (a) and (b) show cases with $Re_{ABC}=894$ for various particle mass fractions. Panels (c) and (d) show cases with $Re_{ABC}=224$ for various particle mass fractions. Panels (e) and (f) show fixed ($M=1$) and single-phase ($M=0$) cases for various Reynolds numbers. {Each case is plotted using its marker, which is listed in table~\ref{tabSinglePhCases}, \ref{tabSphCases}, or \ref{tabFibCases}. The number on the marker shows} the value of $\log p$ at each contour, where $p$ is the joint probability density function. The grey curves show where the discriminant is zero (equation~\ref{eqQR}). The grey shaded regions in the left panels are the loci of equation~\ref{eqQREnhanced}, where we observe an increase $p$ due to the presence of spheres.}
 \label{fig:QR}
\end{figure}

The local topology of a flow can be described entirely using the three principle invariants of the velocity gradient tensor $\partial_i u_j$~\citep{cheng_study_1996}. The first invariant $\partial_i u_i$ is not interesting in our case because it is zero for an incompressible fluid. However, the second invariant, 
\begin{equation}
Q=\frac{1}{4}\omega_i\omega_i-\frac{1}{2}s_{ij}s_{ij},
\end{equation}
expresses the balance of vorticity and strain, while the third invariant,
\begin{equation}
R=\frac{1}{4}\omega_i s_{i j} \omega_j-\frac{1}{3}s_{i j} s_{j k} s_{k i},
\end{equation}
is the balance of the production of vorticity with the production of strain~\cite{paul_role_2022-1}. The eigenvalues $\Lambda$ of $\partial_i u_j$ are the roots of the polynomial equation
\begin{equation}
\Lambda^3+Q \Lambda+R=0. 
\label{eqQR}
\end{equation}
Figure~\ref{fig:QR} shows the joint probability distributions $p(R,Q)$ for all of our flows. We also plot a line where the discriminant of equation~\ref{eqQR} is zero: $27R^2/4+Q^3=0$. Above this line, $\partial_i u_j$ has one real and two complex eigenvalues and vortices dominate the flow. Below this line, all three eigenvalues are real, and the flow is dominated by strain. The single-phase distributions have tails in the top-left and bottom-right quadrants; these correspond to stretching vortices and regions where the flow compresses along one axis, respectively~\citep{cheng_study_1996}. When spheres are added, both $Q$ and $R$ are reduced, as the non-slip boundary condition at their surfaces dampens the flow structures. Incidentally, a similar $Q$ and $R$ reduction has been seen with droplets~\cite{perlekar_kinetic_2019}. However, we see that the addition of spheres causes $p(R,Q)$ to increase in the {shaded} regions {in} the left panels {of} figure~\ref{fig:QR}. {These regions are defined} by 
\begin{equation}
\frac{27R^2}{4}+Q^3<-\frac{4\epsilon^3}{\nu^3}.
\label{eqQREnhanced}
\end{equation}
{These shaded regions} correspond to a strain-dominated flow.
In fact, for pure strain, we have $Q=-s_{ij}s_{ij}/2$ and $R=0$. Hence we can substitute for $Q$ and $R$ in equation~\ref{eqQREnhanced} to estimate that the dissipation is 
\begin{equation}
{2\nu s_{ij}s_{ij} = -4\nu Q < 4^{4/3}\epsilon\approx6\epsilon}
\label{eqSixDiss}
\end{equation}
in the {straining flow. That is, the dissipation in regions around the spheres is roughly six} times greater than the mean dissipation. 
Fibres also create {straining} regions in the flow, seen by the increased probability density functions in the lower parts ($27R^2/4+Q^3<0$) of the right-hand panels in figure~\ref{fig:QR}. As the fibre mass fraction is increased, the variance of $R$ is reduced, showing the production of strain and vorticity is suppressed by the fibres. At high fibre mass fractions the distributions are approximately symmetrical in $R$, in other words, $Q$ and $R$ are uncorrelated (much like the decorrelation of vorticity and strain observed for fibre-laden flows in figure~\ref{fig:align}). This suggests that the small fibre diameter produces many small scale structures which disrupt the typical turbulent flow. 
For both particle types, increasing the Reynolds number $Re_{ABC}$ reduces the effect of the particles. 

\begin{figure}
\includegraphics[width=0.89\textwidth]{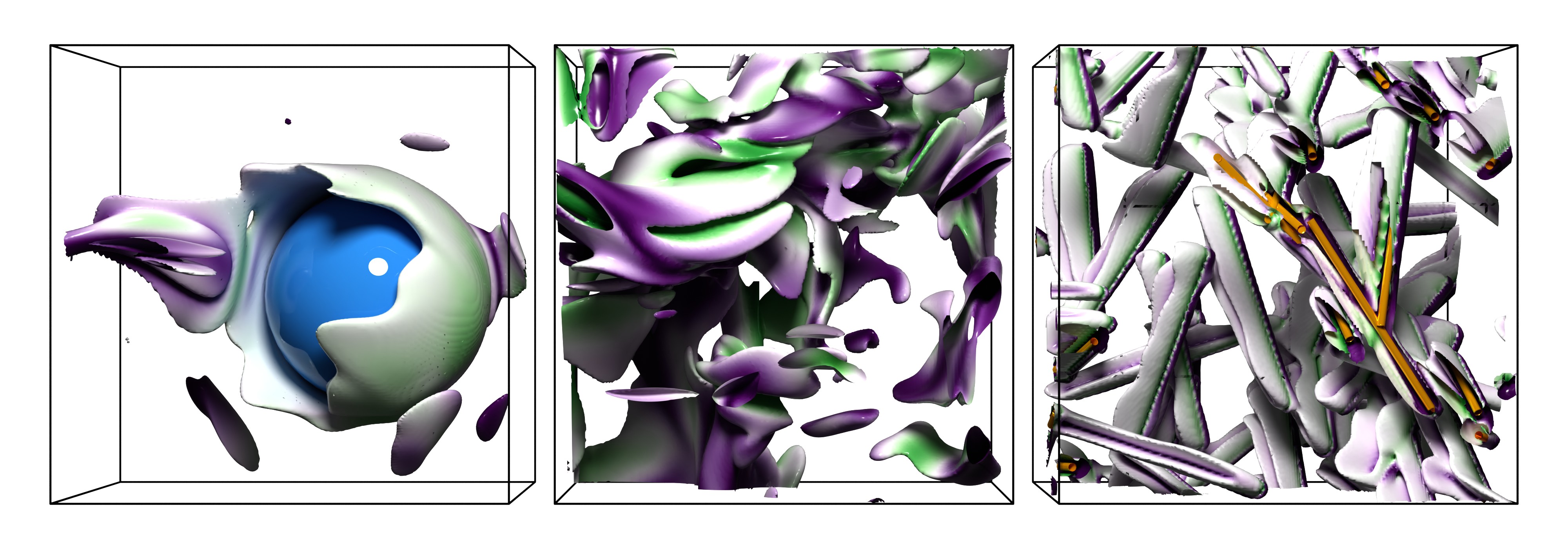}
\includegraphics[width=0.095\textwidth,trim={0 0 0 0}]{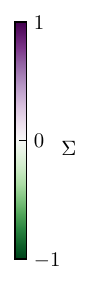}
\caption{{Surfaces where the dissipation is $6\epsilon$, coloured according to the value of the flow topology parameter $\Sigma$, which distinguishes rotational flow (green), shearing flow (white), and straining flow (purple). We show $Re_{ABC}=224$ cases with fixed spheres (left), single-phase (centre), and fixed fibres (right). The boxes have side length $L$.}}
\label{fig:dissSurf}
\end{figure}

{In figure~\ref{fig:dissSurf} we visualise isosurfaces where the local fluid dissipation is six times its bulk value, i.e., where $2\nu s_{ij}s_{ij}=6\epsilon$. As predicted by equation~\ref{eqSixDiss}, the dissipation is greater than $6\epsilon$ in the near-particle regions. And in keeping with our measurements of the fractal dimension of the dissipation field (figure~\ref{fig:fractality}), the high dissipation region near the sphere is approximately sheet-like and two-dimensional, while the high dissipation regions near the fibres are approximately tube-like and one-dimensional. In both cases, wakes extend behind the particles, which contribute to the fractal dimension of the dissipation fields. We see the single-phase dissipation structures take various shapes and sizes, giving rise to the multifractal behaviour discussed above. In figure~\ref{fig:dissSurf} we colour the dissipation isosurfaces according to the local value of the flow topology parameter~\cite{hunt_rapid_1993}, 
\begin{equation}
\Sigma\equiv\frac{2s_{ij}s_{ij}-\omega_i\omega_i}{2s_{ij}s_{ij}+\omega_i\omega_i},
\end{equation}
which is essentially the second invariant of the velocity gradient tensor $Q$ normalised so that it is bounded between~-1 and~1. Regions where $\Sigma=-1$ (coloured green) correspond to pure rotational flow, regions where $\Sigma=0$ (coloured white) correspond to pure shear, and regions where $\Sigma=1$ (coloured purple) correspond to pure straining flow. From figure~\ref{fig:dissSurf} we see fibres produce straining flow on the upstream side, and shearing flow in wakes on the downstream side, this manifests in figure~\ref{fig:QR} as an increase in the spread of $Q$ for the fibre-laden flows. At the surface of the sphere we see a mostly shearing flow, which supports our observations of strong shear-vorticity alignment $(|\boldsymbol{\hat\omega}\cdot\mathbf{\hat a_2}|=1)$ in figure~\ref{fig:align}.}

\section{Conclusion}
\label{sec:concl}
We made simulations of finite-size isotropically- and anisotropically-shaped particles (spheres and fibres with size $c$ within the inertial range of scales) in turbulence at various Reynolds numbers and solid mass fractions. We used bulk flow statistics to show that particles reduce turbulence intensity relative to the single-phase case at all Reynolds numbers, with the fibres producing a more significant reduction effect than the spheres. {Regarding the trend in anomalous} dissipation with Reynolds number, we see that the particle-laden flows {tend to} behave like single-phase flows as $Re_\lambda\to\infty$, and the value of anomalous dissipation is $\epsilon \mathcal{L}/u_{{rms}}^3\approx0.4$ for both single-phase and particle-laden flows. {We see that} spheres slow the convergence rate to the anomalous value, and fibres slow it further. Next, we analysed the flow at each scale of the simulation. Spheres and fibres provide a ``spectral shortcut'' to the flow, removing energy from the largest scales and injecting it at smaller scales. Spheres' effect is mainly limited to the large scales, and they provide a spectral shortcut down to the length scale of their diameter. Fibres' effect, on the other hand, occurs down to the scale of the fibre thickness, even at low Reynolds numbers when it is deep in the viscous range ($d<\eta$). This shortcut of energy to the dissipative scales slows the dissipation's convergence to its anomalous value as $Re_\lambda\to\infty$. Our scale-by-scale analysis also showed that particles cause the velocity field to become more intermittent in space. Multifractal spectra of the near-particle dissipation show that spheres enhance dissipation in two-dimensional sheets, and fibres enhance the dissipation in structures with a dimension greater than one and less than two. These lower dimensional structures are a possible source of intermittency in the flow. Zooming in closer to the flow, we looked at the shape of flow structures using their vorticity and shear. We saw that the particles enhance local shear, spheres suppress vortical flow structures, and fibres produce intense vortical and shearing structures, which overcome the usual vortex stretching behaviour. As Reynolds number increases, the flow structures created by the particles become less significant relative to the background turbulence.

To reiterate and answer our questions from section~\ref{sec:intro}; \emph{at what Reynolds numbers does turbulence modulation emerge?} We see turbulence modulation at the lowest Reynolds number investigated $(Re_\lambda=12.8)$. This gives us reason to believe that particles modulate turbulence in even the most weakly turbulent flows $(Re_\lambda\to0)$. Secondly, \emph{does the modulation effect persist as $Re\to\infty$?} 
{The values of normalised dissipation and many other statistics presented here do indeed converge on the single-phase result as $Re\to\infty$. However, the modulation of turbulent kinetic energy $K'$ appears to have little to no dependence on the Reynolds number. 
Larger Reynolds numbers must be investigated to test whether the attenuation of $K'$ goes to zero as $Re_{ABC}\to\infty$.} Finally, \emph{how do the particle's characteristic lengths impact the scales of the turbulent flow?} Both spheres and {fibres} take energy from the flow at scales larger than their size. Spheres re-inject energy around the characteristic length of their diameter {$c$}, leaving the smaller scales relatively unperturbed, {Indeed, the effect of the spheres on the local flow structures tends to zero as the ratio $c/\eta$ increases. In contrast,} fibres re-inject energy around the characteristic length of their thickness {$d$}. The fibre thickness is at a small scale, so local flow structures are disrupted. None of the flow statistics shows any particular discerning feature at the length scale {$c$} of the fibres. 

These results relate to various environmental particle-laden flows, such as microplastics in the ocean, volcanic ash clouds, and sandstorms. We have explored the two extremes of isotropic and anisotropic particles, but further work is needed to investigate how intermediate aspect ratio particles such as ellipsoids interact with turbulent flows.

\section{Acknowledgements}
The research was supported by the Okinawa Institute of Science and Technology Graduate University (OIST) with subsidy funding from the Cabinet Office, Government of Japan. The authors acknowledge the computer time provided by the Scientific Computing section of Research Support Division at OIST and the computational resources of the supercomputer Fugaku provided by RIKEN through the HPCI System Research Project (Project IDs: hp210229 and hp210269). SO acknowledges the support of grants FJC2021-047652-I and PID2022-142135NA-I00 by MCIN/AEI/10.13039/501100011033 and European Union NextGenerationEU/PRTR.

\section{Data availability}
The data are available at \url{https://groups.oist.jp/cffu/cannon2024PhysRevFluids}.

%\bibliography{particles}
%apsrev4-2.bst 2019-01-14 (MD) hand-edited version of apsrev4-1.bst
%Control: key (0)
%Control: author (8) initials jnrlst
%Control: editor formatted (1) identically to author
%Control: production of article title (0) allowed
%Control: page (0) single
%Control: year (1) truncated
%Control: production of eprint (0) enabled
%

\end{document}